%% file: rfml_concept_bottleneck.tex
\documentclass[conference]{IEEEtran}
\IEEEoverridecommandlockouts
\usepackage{cite}
\usepackage{amsmath,amssymb,amsfonts}
\usepackage{algorithmic}
\usepackage{booktabs}
\usepackage{graphicx}
\usepackage{textcomp}
\usepackage{caption}
\usepackage{subcaption}
\usepackage{acro}
\input{acronyms.tex}
\usepackage[at]{easylist}
\usepackage{xcolor}
\def\BibTeX{{\rm B\kern-.05em{\sc i\kern-.025em b}\kern-.08em
    T\kern-.1667em\lower.7ex\hbox{E}\kern-.125emX}}
\begin{document}

\title{Explainable Neural Network-based Modulation Classification via Concept Bottleneck Models
}

\author{\IEEEauthorblockN{Lauren J. Wong and Sean McPherson}
\IEEEauthorblockA{\textit{Intel AI Lab} \\
lauren.wong@intel.com}
}

\maketitle

\begin{abstract}
While \ac{RFML} is expected to be a key enabler of future wireless standards, a significant challenge to the widespread adoption of \ac{RFML} techniques is the lack of explainability in deep learning models.
This work investigates the use of \ac{CB} models as a means to provide inherent decision explanations in the context of \ac{DL}-based \ac{AMC}.
Results show that the proposed approach not only meets the performance of single-network \ac{DL}-based \ac{AMC} algorithms, but provides the desired model explainability and shows potential for classifying modulation schemes not seen during training (i.e. zero-shot learning).  
\end{abstract}


\section{Introduction}
In recent years, the same \ac{ML} and \ac{DL} technologies that have transformed fields such as \ac{CV} and \ac{NLP} have been extended to the Radio Frequency (RF) domain. 
This field, dubbed \ac{RFML}, has yielded state-of-the-art results in multiple Spectrum Sensing and \ac{CR} applications, including signal detection and classification, \ac{SEI}, and parameter estimation.
All the while, \ac{RFML} has minimized pre-processing, assumed prior knowledge, and end-user domain expertise \cite{darpa_rfmls, wong2020}.
As a result, \ac{RFML} is expected to be a key enabler of future \ac{CR} and \ac{DSA} technologies, and a central component of 5G and 6G standards \cite{morocho2019}.
However, \ac{RFML} algorithms have yet to be regularly integrated into deployed systems and often face resistance due to their ``black box" nature.

Resistance to the widespread adoption of \ac{ML} and \ac{DL} algorithms is not unique to \ac{RFML}, and has yielded a breadth of work in areas such as verification, testing, and interpretation or explanation methods \cite{carvalho2019}. 
The approach presented herein adds to the small body of work in explainable \ac{RFML} methods, providing inherent decision explanations in the context of an 
\ac{AMC} problem in which only the raw IQ data is used as input \cite{dobre2007}.
While it is generally understood that there is a trade-off between model explainability and model performance, this work aims to balance the need for both of these characteristics via \acf{CB} models \cite{koh2020}.

\begin{figure}[t]
	\centering
	\includegraphics[width=\linewidth,trim=0 0 0 0,clip]{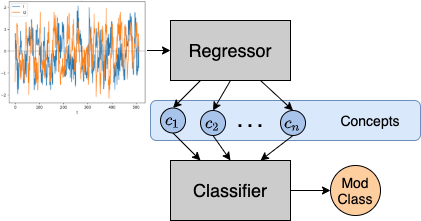}
    \caption{The proposed \ac{CB} model-based approach to Automatic Modulation Classification.}
    \label{fig:approach}
    % \vspace{-0.25cm}
\end{figure}

\ac{CB} models first predict a set of concepts, defined during training, which are then used to make the final prediction.
As a result, \ac{CB} models provide inherent explainability via the predicted concepts, may enable zero-shot learning, as investigated in Section \ref{sec:zeroshot}, and may provide additional metrics for \ac{NN} architecture selection and design.
The proposed \ac{CB} model-based approach, shown in Figure \ref{fig:approach}, provides the desired explainability with little-to-no loss in accuracy compared to a single-network \ac{CNN}-based approach \cite{oshea2016}.

The remainder of this paper is organized as follows: 
First, Section \ref{sec:related} summarizes existing works related to the presented \ac{CB} model-based approach. 
Section \ref{sec:model} describes \ac{CB} models in more detail, including how the models were trained and evaluated. 
Then, Section \ref{sec:dataset} introduces the problem space and discusses the datasets used in the work. 
In Section \ref{sec:results}, results are presented which examine the accuracy of the concept predictions, as well as the final classification, on a number of test sets, and investigate the feasibility of using the proposed approach for zero-shot learning. 
Finally, Section \ref{sec:conclusion} concludes the work, and discusses avenues for future work.

\section{Related Work} \label{sec:related}
\subsubsection{Explainable RFML}
Prior work in explainable \ac{RFML} algorithms primarily utilize ``decision tree" or ``hierarchical" neural networks \cite{clark2019, karra2017, vanhoy2018}, which break the problem space into sub-problems, each solved using separate neural networks.
For example, in \cite{clark2019}, to solve an \ac{AMC} problem, a first network determines the subgroup or modulation type of the input signal (PSK, QAM, FSK, etc.), while networks for each subgroup identify the modulation order, based on the output of the first network.
In practice, networks such as these perform as well as recent single-network \ac{CNN} approaches and provide some explainability, but it should be noted that in the case that the output of the first network is incorrect, the final classification will also necessarily be incorrect. 
Further, such decision-tree approaches tend to be more computationally and memory intensive than the single-network approach, and do not allow for zero-shot learning.

Other works utilize \acp{NN} as naive feature 
learners and extractors in tandem with clustering techniques such as DBSCAN to increase explainability, an approach called ``supervised bootstrapping," for applications such as \ac{AMC} and \ac{SEI} \cite{oshea2017, wong2018}. 
However, while supervised bootstrapping does allow for zero-shot learning, it requires hand-tuning of the clustering parameters and only provides limited explainability in the form of clustering visualizations.

\subsubsection{Multi-task Learning} 
As discussed further in Section \ref{sec:model}, \ac{CB} models are composed of a multi-head regression model, which performs the concept predictions, and a classifier, which utilizes the concept predictions as input to make the final classification.
As a result, the proposed \ac{CB} model-based approach can be viewed as a multi-task learning problem, an approach often utilized in the context of transfer learning. 
Other prior works in the realm of \ac{RFML} have also utilized multi-task learning techniques, including the aforementioned ``hierarchichal" neural networks \cite{clark2019}, as well as for performing non-orthogonal multiple access \cite{gui2018}.

\section{Concept Bottleneck Models} \label{sec:model}
\acf{CB} models first predict a number of concepts, provided during training, then use these predicted concepts to predict the target label.
The \ac{CB} model itself is composed of two networks: a multi-head regression network, $\hat{c} = g(x)$, which estimates the pre-defined concepts, $c$, using $x$ as input, and classification network, $\hat{y} = f(\hat{c})$, which predicts the target label, $y$, using only the estimated concepts, $\hat{c}$, as input.
As a result, the regressor acts as a ``bottleneck" in data flow. For the given \ac{AMC} problem, $x$ is the raw IQ data in the form of a $2 \times 128$ vector, and $y$ is the modulation scheme.

\subsubsection{Learning Concept Bottleneck Models}
As in the original \ac{CB} work \cite{koh2020}, three training regimes are examined in the results below: Independent, Sequential, and Joint. A standard trained network is included as a baseline. 

\begin{enumerate}
    \item The \textit{independent bottleneck} learns $\hat{y} = \hat{f}(c)$ and $\hat{c} = \hat{g}(x)$ separately.
    During testing, $\hat{y} = \hat{f}(\hat{g}(x))$.
    \item The \textit{sequential bottleneck} learns $\hat{c} = \hat{g}(x)$ first, then learning $\hat{y} = \hat{f}(\hat{c})$.
    \item The \textit{joint bottleneck} trains both networks at once, learning $\hat{y} = \hat{f}(\hat{g}(x))$.
    \item The \textit{standard model} does not learn concepts, and rather learns $\hat{y} = \hat{f}(x)$.
\end{enumerate}
Of the \ac{CB} regimes, the \textit{independent bottleneck} method is most efficient as the regressor and classifier can be trained in parallel. However, any error in the regressor is not considered in training the classifier, as the classifier is trained using the pristine concepts, $c$, as input, rather than those predicted by the regressor, $\hat{c}$. The \textit{sequential bottleneck} method rectifies this shortcoming by training the classifier with concepts predicted by the regressor, $\hat{c}$, however, the two networks are trained with two independent loss functions. The \textit{joint bottleneck} is the most complex method, as it utilizes a single loss function for training both networks. Specifically, this approach utilizes a weighted loss term, and therefore introduces an additional hyper-parameter which specifies the weight of the classifier's loss term in relation to the regressor's loss terms.

\subsubsection{Concept Selection}
The choice of concepts making up $c$ is largely a design choice, and is task specific.
However, it should be noted that the chosen concepts must completely describe the label space, as no auxiliary data can flow through the network in it's current form \cite{bahadori2020}.
In this work, $c$ is a vector containing the following five values, selected to provide maximal explainability:
%SM: if you hate the math equations we can throw them out! 
\begin{itemize}
    \item $c_M= I_M(x):=\begin{cases}
                          $1$ & \text{if $x$ is Analog,}\\
                          $0$ & \text{if $x$ is Digital.}
                        \end{cases}$
    \item $c_A= I_A(x):=\begin{cases}
                          $1$ & \text{if $x$ is Amplitude Modulated,}\\
                          $0$ & \text{Otherwise.}
                        \end{cases}$
    \item $c_P= I_P(x):=\begin{cases}
                          $1$ & \text{if $x$ is Phase Modulated,}\\
                          $0$ & \text{Otherwise.}
                        \end{cases}$
    \item $c_F= I_F(x):=\begin{cases}
                          $1$ & \text{if $x$ is Frequency Modulated,}\\
                          $0$ & \text{Otherwise.}
                        \end{cases}$
    \item $c_O= \begin{cases}
                     \frac{1}{\log_2(order(x))} & \text{if $order(x)$ is Defined,}\\
                     $\qquad 0$ & \text{if $order(x)$ is Undefined. }
                    \end{cases}$
\end{itemize}

\begin{table}[]
\centering
\caption{Regressor architecture.}
\label{tab:model_regressor}
\resizebox{0.45\textwidth}{!}{%
\begin{tabular}{@{}lccc@{}}
\toprule
\multicolumn{1}{c}{Layer Type} & Num Kernels/Nodes & Kernel Size & Padding \\ \midrule
Input & size = (2, 128) &  &  \\
Conv2d & 96 & (1, 21) & (0, 10) \\
ReLU &  &  &  \\
Dropout & rate = 0.5 &  &  \\
Conv2d & 96 & (2, 21) & (0, 10) \\
ReLU &  &  &  \\
Dropout & rate = 0.5 &  &  \\
Flatten &  &  &  \\
Linear & 384 &  &  \\
ReLU &  &  &  \\
Dropout & rate = 0.5 &  &  \\
Linear & 5 &  &  \\ \midrule
\multicolumn{4}{c}{Trainable Parameters: 9830313} \\ \bottomrule
\end{tabular}%
}
% \vspace{-0.25cm}
\end{table}

\begin{table}[]
\centering
\caption{Classifier architecture.}
\label{tab:model_classifier}
\resizebox{0.22\textwidth}{!}{%
\begin{tabular}{@{}lc@{}}
\toprule
\multicolumn{1}{c}{Layer Type} & Num Nodes \\ \midrule
Input & 5 \\
Linear & 64 \\
ReLU & \multicolumn{1}{l}{} \\
Dropout & rate = 0.5 \\
Linear & 64 \\
ReLU & \multicolumn{1}{l}{} \\
Dropout & rate = 0.5 \\
Dense & 9 \\ \midrule
\multicolumn{2}{c}{Trainable Parameters: 5129} \\ \bottomrule
\end{tabular}%
}
\end{table}

% \begin{table}[]
% \centering
% \caption{Baseline CNN architecture.}
% \label{tab:model_oshea}
% \resizebox{0.45\textwidth}{!}{%
% \begin{tabular}{@{}lccc@{}}
% \toprule
% \multicolumn{1}{c}{Layer Type} & Num Kernels/Nodes & Kernel Size & Padding \\ \midrule
% Input & size = (2, 128) &  &  \\
% Conv2d & 96 & (1, 21) & (0, 10) \\
% ReLU & \multicolumn{1}{l}{} & \multicolumn{1}{l}{} & \multicolumn{1}{l}{} \\
% Dropout & rate = 0.5 &  &  \\
% Conv2d & 96 & (2, 21) & (0, 10) \\
% ReLU &  &  &  \\
% Dropout & rate = 0.5 &  &  \\
% Flatten &  &  &  \\
% Linear & 384 &  &  \\
% ReLU & \multicolumn{1}{l}{} & \multicolumn{1}{l}{} & \multicolumn{1}{l}{} \\
% Dropout & rate = 0.5 &  &  \\
% Linear & 9 &  &  \\ \midrule
% \multicolumn{4}{c}{Trainable Parameters: 9830313} \\ \bottomrule
% \end{tabular}%
% }
% % \vspace{-0.25cm}
% \end{table}

\subsubsection{Model Description}
In this work, the concept regressor is a \ac{CNN} with five regression heads, each using a \ac{MSE} loss function, equally weighted during training.
The classifier is an \ac{MLP} and uses a Categorical Cross Entropy loss function.
These architectures were chosen using a number of hyper-parameter sweeps.
The baseline \ac{CNN} simply is a slightly wider version of that used in \cite{oshea2016}, and is identical to the regression architecture, with a softmax output layer instead the 5 regression heads.
The specifications for each of these networks is given in Tables \ref{tab:model_regressor} and \ref{tab:model_classifier}.

Additionally, for the models trained using the joint bottleneck training regime, the classifier's loss makes up 0.3 of the total loss, with the remaining 0.7 equally distributed amongst the regression heads.
As a result, the classifier's loss is weighed approximated double that of any single regression head in the total loss term.
This hyper-parameter was chosen using additional hyper-parameter sweeps.

All models used in this work were trained using the Adam optimizer with a learning rate of 0.0001.
Each regressor was trained for 200 epochs and each classifier was trained for 100 epochs, both with checkpoints saved at epochs with the lowest validation loss.

\section{Problem Space and Dataset} \label{sec:dataset}
\begin{table}[t]
\centering
\caption{In-Set Signals and Generation Parameters}
\label{tab:inset}
\resizebox{0.45\textwidth}{!}{%
\begin{tabular}{@{}ccl@{}}
\toprule
\begin{tabular}[c]{@{}c@{}}Modulation\\ Name\end{tabular} & \begin{tabular}[c]{@{}c@{}}Symbol\\ Order\end{tabular} & \multicolumn{1}{c}{\begin{tabular}[c]{@{}c@{}}Parameter \\ Space\end{tabular}} \\ \midrule
BPSK \vspace{0.25cm} & 2 & \begin{tabular}[c]{@{}l@{}}RRC Pulse Shape\\ Excess Bandwidth \{0.35, 0.5\}\\ Symbol Overlap \{3, 5\}\end{tabular} \\
QPSK \vspace{0.25cm} & 4 & \begin{tabular}[c]{@{}l@{}}RRC Pulse Shape\\ Excess Bandwidth \{0.35, 0.5\}\\ Symbol Overlap \{3, 5\}\end{tabular} \\
8PSK \vspace{0.25cm} & 8 & \begin{tabular}[c]{@{}l@{}}RRC Pulse Shape\\ Excess Bandwidth \{0.35, 0.5\}\\ Symbol Overlap \{3, 5\}\end{tabular} \\
16QAM \vspace{0.25cm} & 16 & \begin{tabular}[c]{@{}l@{}}RRC Pulse Shape\\ Excess Bandwidth \{0.35, 0.5\}\\ Symbol Overlap \{3, 5\}\end{tabular} \\
64QAM \vspace{0.25cm} & 64 & \begin{tabular}[c]{@{}l@{}}RRC Pulse Shape\\ Excess Bandwidth \{0.35, 0.5\}\\ Symbol Overlap \{3, 5\}\end{tabular} \\
FSK \vspace{0.25cm} & 2 & \begin{tabular}[c]{@{}l@{}}Rect Phase Shape\\ Carrier Spacing 5kHz, 75kHz\end{tabular} \\
AM \vspace{0.25cm} & - & \begin{tabular}[c]{@{}l@{}}DSB Mode\\ Modulation Index {[}0.5, 0.9{]}\end{tabular} \\
FM-NB \vspace{0.25cm} & - & Modulation Index {[}0.05, 0.4{]} \\
AWGN & - &  \\ \bottomrule
\end{tabular}%
}
\end{table}

\begin{table}[t]
\centering
\caption{Out-of-Set Signals and Generation Parameters}
\label{tab:outset}
\resizebox{0.45\textwidth}{!}{%
\begin{tabular}{@{}ccl@{}}
\toprule
\begin{tabular}[c]{@{}c@{}}Modulation\\ Name\end{tabular} & \begin{tabular}[c]{@{}c@{}}Symbol\\ Order\end{tabular} & \multicolumn{1}{c}{\begin{tabular}[c]{@{}c@{}}Parameter \\ Space\end{tabular}} \\ \midrule
16PSK \vspace{0.25cm} & 16 & \begin{tabular}[c]{@{}l@{}}RRC Pulse Shape\\ Excess Bandwidth \{0.35, 0.5\}\\ Symbol Overlap \{3, 5\}\end{tabular} \\
32QAM \vspace{0.25cm} & 32 & \begin{tabular}[c]{@{}l@{}}RRC Pulse Shape\\ Excess Bandwidth \{0.35, 0.5\}\\ Symbol Overlap \{3, 5\}\end{tabular} \\
MSK \vspace{0.25cm} & 2 & \begin{tabular}[c]{@{}l@{}}Rect Phase Shape\\ Carrier Spacing 2.5kHz \end{tabular} \\ 
GFSK \vspace{0.25cm} & 2 & \begin{tabular}[c]{@{}l@{}}Gaussian Phase Shape\\ Carrier Spacing 5kHz, 75kHz\\Symbol Overlap \{2, 4\}\\Beta \{0.3, 0.5\}\end{tabular} \\ 
AM \vspace{0.25cm} & - & \begin{tabular}[c]{@{}l@{}}LSB Mode\\ Modulation Index {[}0.5, 0.9{]}\end{tabular} \\ 
FM-WB & - & Modulation Index {[}0.825, 1.88{]}\\
\bottomrule
\end{tabular}%
}
\end{table}

For this work, all signals which compose the training, validation, and test sets are observed at complex baseband, as given by 
$$ s[t] = \alpha_\Delta[t] \cdot \alpha[t]e^{j\omega[t] + j\theta[t]} \cdot e^{(j\omega_\Delta[t] + j\theta_\Delta[t])} + \nu[t] $$
where $\alpha[t]$ represents instantaneous magnitude, $\omega[t]$ represents instantaneous frequency, and $\theta[t]$ represents instantaneous phase, at time $t$.
Frequency offsets are selected uniformly at random on the interval $[-0.1\pi, 0.1\pi]$ and are held constant for each signal realization (i.e. $\omega_\Delta[t] = U(-0.1\pi, 0.1\pi))$, and no phase offset is applied (i.e. $\theta_\Delta[t] = 0$).
All signals are sampled at either 4 or 8 samples per symbol, or 2 or 4 times Nyquist.

Additionally, all signals are observed in an \ac{AWGN} environment with unit channel gain (i.e. $\alpha_\Delta[t]=1$), where $\nu[t]$ describes the additive interference or noise at time $t$.
For each capture, the value of $\nu[t]$ is defined such that the \ac{SNR}, defined as
$$
\Gamma_s = 10 \log_{10} \bigg( \frac{\sum_{t=0}^{N-1} \lvert s[t] - \nu[t] \rvert^2}
{\sum_{t=0}^{N-1}\lvert \nu[t] \rvert^2} \bigg)
$$
is an integer value selected uniformly at random on the interval $[0, 20]$dB for a capture containing $N$ samples.

The problem space is further constrained to the 15 modulation schemes shown in Tables \ref{tab:inset} and \ref{tab:outset}. From these modulation schemes, three categories of signals are defined:
\begin{itemize}
    \item \textit{in-set}: includes modulation schemes given in Table \ref{tab:inset}, observed during training, validation, and testing. 
    \item \textit{near-set}: includes modulation schemes in Table \ref{tab:inset}, observed during testing with \acp{SNR} selected uniformly at random on the intervals $[-5, -1]$dB and $[21, 25]$dB. Used to measure the models ability to generalize. 
    \item \textit{out-of-set}: includes modulation schemes show in Table \ref{tab:outset}, which are only seen during testing, with the parameters as given, to gauge the zero-shot learning capabilities of the \ac{CB} approach.
\end{itemize}

This work utilizes training, validation, and testing composed of synthetic signals generated using a custom Python wrapper around \textit{liquid-dsp} \cite{gaeddert-liquid}.
The training set contained 10000 examples per class (90000 total), and the validation set contained 5000 examples per class (45000 total).
Additionally, three test sets were used to evaluate the performance of the approach on in-set signals, near-set signals, and out-of-set signals, each of which contained 10000 examples per class (90000, 80000, and 60000 total). 

\section{Results}\label{sec:results}
As previously described in Section \ref{sec:model}, one regressor and classifier architecture is evaluated across the three \ac{CB} training methods (Independent, Sequential, and Joint) and the three test datasets (In-Set, Near-Set, and Out-of-Set).
In the following subsections, the performance of these models is compared to each other to identify the strengths and weaknesses of each training method, as well as to a baseline CNN architecture. 

\subsection{Classifier Performance}\label{sec:classifier_results}

\begin{figure}[t]
	\centering
	\includegraphics[width=\linewidth,trim=0 0 0 0,clip]{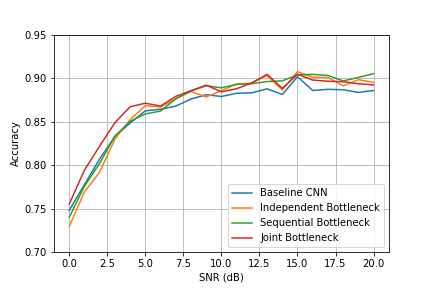}
    \caption{In-set accuracy versus \ac{SNR} for all three concept bottleneck models and the baseline \ac{CNN} model.}
    \label{fig:accVsnr}
\end{figure}

%SM: assuming we get the figures looking better, do you like this version? 
\begin{figure*}
    \centering
    \begin{subfigure}[t]{0.22\textwidth}
        \centering
        \includegraphics[width=1.5\textwidth,trim=10 0 0 0,clip]{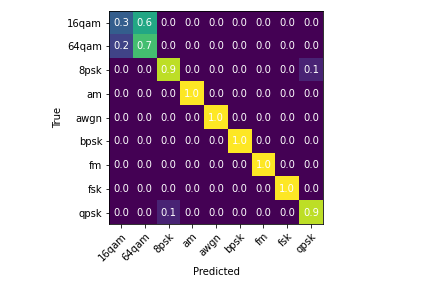}
    	\caption{Independent}
        \label{fig:ind_cm}
    \end{subfigure}
    % \hfill
    \begin{subfigure}[t]{0.22\textwidth}
    	\centering
        \includegraphics[width=1.5\textwidth,trim=10 0 0 0,clip]{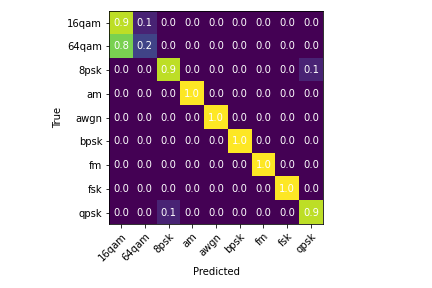}
    	\caption{Sequential}
        \label{fig:seq_cm}
    \end{subfigure}    
    % \hfill
    \begin{subfigure}[t]{0.22\textwidth}
	    \centering
	   \includegraphics[width=1.5\textwidth,trim=10 0 0 0,clip]{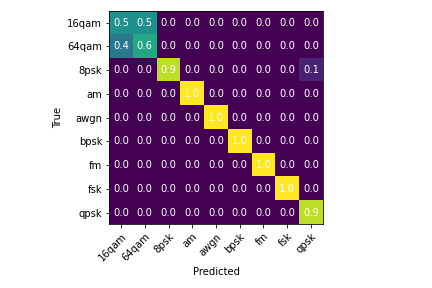}
	    \caption{Joint}
        \label{fig:joint_cm}
    \end{subfigure}
    % \hfill
    \begin{subfigure}[t]{0.3\textwidth}
	    \centering
	   \includegraphics[width=1.04\textwidth,trim=30 0 10 0,clip]{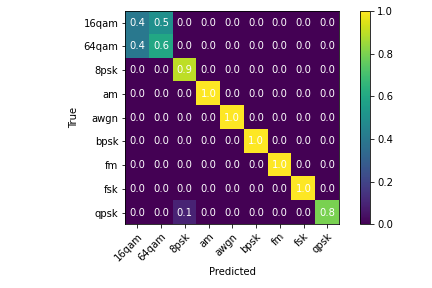}
	    \caption{Baseline CNN}
        \label{fig:cnn_cm}
    \end{subfigure}
    \caption{Confusion matrices for each learning scheme on the \textit{in-set} test data. }
    \label{fig:cm}
    \vspace{-0.25cm}
\end{figure*}

Considering first the in-set classification accuracy of each \ac{CB} model in comparison to the baseline CNN approach in Table \ref{tab:accuracy} and Figure \ref{fig:accVsnr}.
Both of these figures indicate that the \ac{CB} models marginally outperform the baseline CNN approach on the in-set test dataset. 
Furthermore, all three \ac{CB} models achieve overall in-set accuracies within 1\% of each other, with the Joint model achieving the highest in-set classification accuracy of 87.24\%.
Additionally, Figure \ref{fig:cm} shows that the primary source of confusion for all models, including both the \ac{CB} models and baseline CNN, is 16QAM and 64QAM, which is to be expected, particularly at low \acp{SNR}.

\begin{table}[t]
\centering
\caption{In-set and near-set classification accuracy for all models, averaged over all \acp{SNR}.}
\label{tab:accuracy}
\resizebox{0.4\textwidth}{!}{%
\begin{tabular}{@{}lcc@{}}
\toprule
\multicolumn{1}{c}{Model} & \begin{tabular}[c]{@{}c@{}}In-Set\\ Accuracy\end{tabular} & \begin{tabular}[c]{@{}c@{}}Near-Set \\ Accuracy\end{tabular} \\ \midrule
Baseline CNN & 0.8622 & 0.6836 \\
Independent Bottleneck & 0.8667 & 0.7151 \\
Sequential Bottleneck & 0.8693 & 0.7037 \\
Joint Bottleneck & 0.8724 & 0.6880 \\ \bottomrule
\end{tabular}%
}
\end{table}

Table \ref{tab:accuracy} also shows that all three \ac{CB} models outperform the baseline \ac{CNN} approach on the near-set test dataset.
However, while the Joint model outperformed the Independent and Sequential models on the in-set test data, the opposite is true in the case of the near-set test data.
That is, the Independent model, which was the lowest performing amongst the \ac{CB} models on the in-set data, is the highest performing on the near-set test data.

\subsection{Regressor Performance}\label{sec:regressor_results}

\begin{figure*}
	\centering
	\begin{subfigure}[t]{.48\textwidth}

	    \includegraphics[width=\linewidth,trim=0 0 0 0,clip]{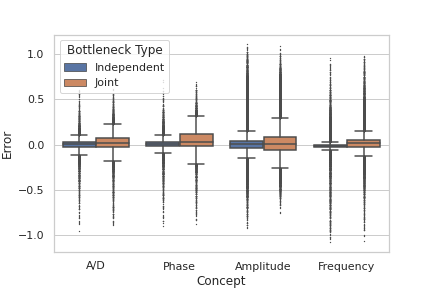}
        \caption{Binary regression heads.}
        \label{fig:regressor_binary}
    \end{subfigure}
    \begin{subfigure}[t]{.48\textwidth}
	    \centering
	    \includegraphics[width=\linewidth,trim=10 0 0 0,clip]{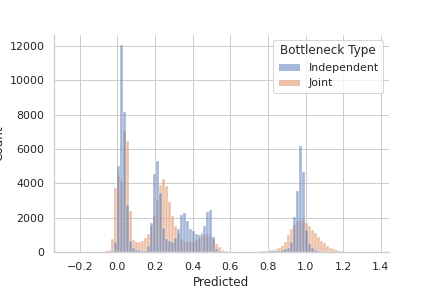}
        \caption{Order regression head.}
        \label{fig:regressor_order}
    \end{subfigure}
    \caption{Performance of regression heads on \textit{in-set} data. }
    \label{fig:regressor_perf}
    % \vspace{-0.25cm}
\end{figure*}

One of the primary benefits of the proposed \ac{CB} model-based approach over the baseline \ac{CNN}-based approach is the availability of decision explanations in the form of the intermediate predicted concepts.
The accuracy of these predicted concepts on the in-set test dataset is shown in Figure \ref{fig:regressor_perf}.
More specifically, for the Independent Bottleneck and Joint Bottleneck models, Figure \ref{fig:regressor_binary} plots the error in the regression heads with binary truth values (A/D, Phase, Amplitude, and Frequency) in the form of a box-plot and Figure \ref{fig:regressor_order} plots the predicted order (more specifically the concept, $\hat{c}_O$) in the form of a histogram.
Note that the Independent and Sequential models differ only in the classifier models, and utilize the same regressor models. 
Therefore, results in this section are shown for the Independent and Joint Bottleneck regressors only.

Figure \ref{fig:regressor_binary} shows that both the Independent and Joint models predict the binary concepts with low error on in-set data.
However, the Independent model shows stronger overall regression performance than the Joint Bottleneck model, despite the fact that the Joint Bottleneck outperforms the Independent and Sequential Bottleneck models in terms of overall accuracy.
This result is evident in Figure \ref{fig:regressor_order} as well, where the histogram of the Independent Bottleneck model order regression-head contains all five peaks, while the histogram of the Joint Bottleneck model order regression-head only contains four peaks and has created a single output range for the higher order modulation schemes: 16 and 64.

These trends show that the Joint training regime results in marginally improved overall accuracy at the cost of losing some model explainability, as the accuracy of the concept predictions decreases.
In addition to providing more transparent models, the Independent and Sequential training regimes also result in models more robust to near-set input.
Therefore, the choice of training regime is dependent upon the relative importance of explainablity and classification accuracy to the designer/end-user.
More specifically, the Independent model is the most explainable and robust to near-set input, the Joint model provides the highest in-set classification accuracy, and the Sequential model provides a sort of middle ground with quality decision explanations and good in-set and near-set classification accuracy.

\subsection{Zero-shot Learning}\label{sec:zeroshot}

Another significant benefit of using the \ac{CB} model-based approach over the baseline \ac{CNN}-based approach is the potential for zero-shot learning, or the ability to classify signals outside of the training distribution, whether that be near-set or out-of-set (as described in Section \ref{sec:dataset}).
While the classification accuracy of the proposed \ac{CB} approach on the near-set test dataset was discussed previously, this section examines the performance of the \ac{CB} regression models on both the near-set and out-of-set test datasets.
More accurate regression models, especially on the near-set and out-of-set test datasets, points to better zero-shot learning performance.
To implement zero-shot learning in the proposed \ac{CB} framework, the classifier component of the \ac{CB} model would need to be re-trained, augmenting the training dataset with examples, ($c_R$, $y_R$), where $c_R$ are the concepts and $y_R$ are the target labels for the new signals of interest.
However, this is left for future work.

\begin{figure}[t]
	\centering
	\includegraphics[width=\linewidth,trim=0 0 0 0,clip]{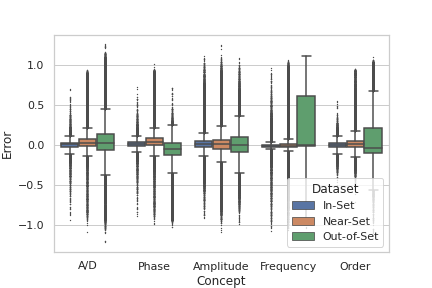}
    \caption{Regression performance across the in-set, near-set, and out-of-set data types for the Independent Bottleneck model.}
    \label{fig:regressor_datasets}
    % \vspace{-0.25cm}
\end{figure}

Figure \ref{fig:regressor_datasets} compares the prediction error across the three test datasets, and shows similar regression performance between the in-set and near-set datasets, which echos the classification results shown in Table \ref{tab:accuracy}.
However, while the A/D, Phase, and Amplitude binary regression heads continue to perform well on the out-of-set test dataset, Frequency and order prediction performance declines.

\begin{figure}[t]
	\centering
	\vspace{.55cm}
	\includegraphics[width=\linewidth,trim=0 0 0 0,clip]{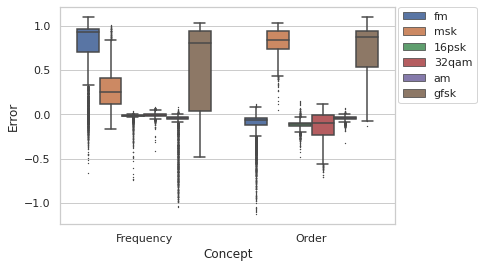}
    \caption{Out-of-set frequency and order prediction using the Independent Bottleneck model broken out by modulation scheme.}
    \label{fig:regressor_outset}
    % \vspace{-0.25cm}
\end{figure}

Figure \ref{fig:regressor_outset} expands upon the results given in figure \ref{fig:regressor_datasets} by breaking out the performance of the Frequency and order regression heads by modulation scheme on the out-of-set signal types, and shows that for these more challenging concepts, FM-WB, GFSK, and MSK are most challenging for the regression model to accurately assess.
Intuitively, these modulation schemes differ most from the in-set modulation schemes, in comparison to the remaining out-of-set modulation schemes, namely 16PSK, 32QAM, and AM-LSB.
This result echos those shown in the zero-shot learning literature which indicate that algorithms similar to the \ac{CB} approach often suffer when the source data and target data are disjoint 
\cite{fu2015}.
Therefore, \ac{CB} model-based zero-shot learning is likely feasible for near-set signal types and signal types which are similar to those seen in training.
However, further work is likely needed to improve the accuracy of the regression model to improve zero-shot learning performance on out-of-set signal types.

\section{Conclusion}\label{sec:conclusion}

In this work, \acf{CB} models were investigated as a means to provide decision explanations in the context of a \acf{DL}-based \acf{AMC} algorithm.
The approach provides inherent decision explanations, in the form of intermediate concept predictions, in addition to the predicted modulation class, using only raw IQ data as input.

The approach was evaluated across three training regimes and three test sets containing in-set, near-set, and out-of-set data.
Results show the \ac{CB} models achieve similar classification accuracy to a recent single-network \ac{AMC} algorithm on in-set data, and outperforms said algorithm on the near-set test dataset.
Further, the proposed approach shows promise evaluating signal types unseen in training (out-of-set data) using the intermediate predicted concepts, a setting known as zero-shot learning.

Most apparently, future work includes applying the \ac{CB} to other \acf{RFML} applications, and improving accuracy through the use of more sophisticated architecures, loss functions, and hyper-parameter tuning.
In particular, the regression component of the \ac{CB} model needs improvement if zero-shot learning is desired, another avenue for future work.
The accuracy of the intermediate concept predictions could also be investigated as another set of metrics for \ac{NN} architecture selection.  
Finally, a more intriguing avenue for future work includes \textit{debiasing} the model using a causal prior graph, to eliminate correlations between the intermediate concepts and spurious features such as \ac{SNR} \cite{bahadori2020}.

% However, improvements could be made 

% \section*{Acknowledgment}

\bibliographystyle{IEEEtran}
\bibliography{rfml_concept_bottleneck}

\end{document}

%% file: acronyms.tex
\DeclareAcronym{AMC}{
  short = AMC ,
  long  = Automatic Modulation Classification ,
  tag = abbrev
}

\DeclareAcronym{DL}{
  short = DL ,
  long  = deep learning ,
  tag = abbrev
}

\DeclareAcronym{SEI}{
    short = SEI ,
    long = Specific Emitter Identification ,
    tag = abbrev
}

\DeclareAcronym{CNN}{
    short = CNN ,
    long = Convolutional Neural Network ,
    tag = abbrev
}

\DeclareAcronym{NLP}{
  short = NLP ,
  long  = Natural Language Processing ,
  tag = abbrev
}

\DeclareAcronym{RFML}{
  short = RFML ,
  long  = Radio Frequency Machine Learning ,
  tag = abbrev
}

\DeclareAcronym{DARPA}{
  short = DARPA ,
  long  = Defense Advanced Research Projects Agency ,
  tag = abbrev
}

\DeclareAcronym{IQ}{
  short = IQ ,
  long  = In-Phase and Quadrature ,
  tag = abbrev
}

\DeclareAcronym{DSA}{
  short = DSA ,
  long  = Dynamic Spectrum Access ,
  tag = abbrev
}

\DeclareAcronym{DNN}{
  short = DNN ,
  long  = Deep Neural Network ,
  tag = abbrev
}

\DeclareAcronym{NN}{
  short = NN ,
  long  = Neural Network ,
  tag = abbrev
}

\DeclareAcronym{CV}{
  short = CV ,
  long  = Computer Vision ,
  tag = abbrev
}

\DeclareAcronym{CR}{
  short = CR ,
  long  = Cognitive Radio ,
  tag = abbrev
}

\DeclareAcronym{RF}{
  short = RF ,
  long  = Radio Frequency ,
  tag = abbrev
}

\DeclareAcronym{AWGN}{
  short = AWGN ,
  long  = Additive White Gaussian Noise ,
  tag = abbrev
}

\DeclareAcronym{SNR}{
  short = SNR ,
  long  = Signal-to-Noise Ratio ,
  tag = abbrev
}

\DeclareAcronym{ML}{
  short = ML ,
  long  = Machine Learning ,
  tag = abbrev
}

\DeclareAcronym{MLP}{
  short = MLP ,
  long  = Multi-Layer Perceptron ,
  tag = abbrev
}

\DeclareAcronym{MSE}{
  short = MSE ,
  long  = Mean Squared Error ,
  tag = abbrev
}

\DeclareAcronym{CB}{
  short = CB ,
  long  = Concept Bottleneck ,
  tag = abbrev
}

\DeclareAcronym{XAI}{
  short = XAI ,
  long  = Explainable Artificial Intelligence ,
  tag = abbrev
}

\DeclareAcronym{QAM}{
  short = QAM ,
  long  = Quadrature Amplitude Modulation ,
  tag = abbrev
}

\DeclareAcronym{PSK}{
  short = PSK ,
  long  = Phase Shift Keying ,
  tag = abbrev
}

\DeclareAcronym{FSK}{
  short = FSK ,
  long  = Frequency Shift Keying ,
  tag = abbrev
}

\DeclareAcronym{CFO}{
  short = CFO ,
  long  = Center Frequency Offset ,
  tag = abbrev
}